\documentclass[aps,prl,twocolumn,amsfonts,showpacs]{revtex4}
\usepackage{bm}
\usepackage{epsfig,amsopn}
\usepackage{graphicx}
\usepackage{amsmath,amssymb}
\usepackage{amsthm}
\usepackage{natbib}
\newcommand\bea{\begin{eqnarray}}
\newcommand\eea{\end{eqnarray}}
\newcommand\beq{\begin{equation}}
\newcommand\eeq{\end{equation}}

\newcommand{\bib}{\bibitem}

\def\nn{\nonumber}
\def\dg{\dagger}
\def\f{\frac}
\def\la{\langle}
\def\ra{\rangle}

\def\b{\beta}

\def\e{\epsilon}
\def\g{\gamma}

\def\ua{\uparrow}
\def\da{\downarrow}
\def\bk{{\bf k}}
\def\bkn{{\bf k}_{N}}
\def\bq{{\bf q}}
\def\bqn{{\bf q}_{N}}
\def\bx{{\bf x}}
\def\b0{{\bf 0}}
\def\hj{{\hat j}_x}
\def\dj{{\delta j}}
\def\noi{\noindent}

\begin{document}

\title{Nonequilibrium charge transport in an interacting open system: 
two-particle resonance and current asymmetry}
\author{Dibyendu Roy,$^1$ Abhiram Soori,$^2$ Diptiman Sen,$^2$ and Abhishek 
Dhar$^1$}
\affiliation{$^1$Raman Research Institute, Bangalore 560080, India \\
$^2$ Centre for High Energy Physics, Indian Institute of Science,
Bangalore 560012, India}

\begin{abstract}
We use Lippman-Schwinger scattering theory to study nonequilibrium 
electron transport through an interacting open quantum dot. 
The two-particle current is evaluated exactly while we use perturbation 
theory to calculate the current when the leads are Fermi liquids at different 
chemical potentials. We find an interesting two-particle resonance induced by 
the interaction and obtain criteria to observe it when a small bias is applied
across the dot. Finally, for a system without spatial inversion symmetry 
we find that the two-particle current is quite different depending on whether 
the electrons are incident from the left lead or the right lead.
\end{abstract}
\vskip .5 true cm

\maketitle

We study nonequilibrium steady state charge transport in an open quantum 
system in the presence of a repulsive Coulomb interaction in a localized 
region. One of the simplest realizations of our model is a quantum 
dot (QD) connected to two noninteracting leads at different chemical 
potentials. In the last two decades, there have been several theoretical 
\cite{todorov,ng,wing,oreg,aharony,mehta,nishino,goorden,lebedev,dhar,boulat}
and experimental 
\cite{ralph,mcclure,hubel,goldhaber,cronenwett,wiel,leturcq,potok} studies 
of electron transport through a QD where electrons interact with each other 
only in the dot region. The presence of a chemical potential difference 
across the QD leads to nonequilibrium dynamics which opens up the possibility 
of exploring the interplay of nonequilibrium physics and interactions in this 
model. In this spirit, we will study two interesting phenomena in our model 
system, namely, two-particle resonance and current asymmetry.

The phenomenon of resonances is often realized in open quantum systems. 
Resonances are signatures of quasi-stationary states with a long life-time 
which eventually decay into the continuum coupled to them. There are many 
examples of resonances in different branches of physics, especially atomic and
nuclear physics. Systems with or without interactions between the constituents
like electrons, photons or phonons can exhibit resonances; for example, the 
symmetric Breit-Wigner \cite{mello} or the asymmetric Fano resonances 
\cite{fano} can occur in noninteracting systems, while the Kondo resonance 
\cite{goldhaber,cronenwett,wiel,leturcq,potok} occurs in correlated electronic
systems. In a recent work \cite{shen}, strongly correlated two-photon 
transport in a one-dimensional system was studied. In this paper, we study 
a two-electron resonance which occurs due to the interactions between 
electrons; this was recently observed in Ref. \cite{lebedev}. This resonance 
is clearly visible in the two-electron current. We demonstrate that it 
survives in the thermodynamic limit when one takes the leads to be 
Fermi seas of electrons. Our two-electron resonance can occur at small bias 
and when the one-particle current is small; it differs from the 
pair-tunneling resonance studied in Ref. \cite{leijnse} which requires a 
sufficiently large bias between the leads and coexists with one-particle 
transport.

A rectification of the current can be achieved in a system without spatial 
inversion symmetry. There are many theoretical and experimental studies of 
the diode effect in electron transport using the nonlinear regime of transport
in asymmetric nanostructures \cite{song}, Coulomb blockade in triple QD 
\cite{stopa} or Pauli exclusion in coupled double QD \cite{ono}. Current 
rectification has also been realized in thermal and optical systems 
\cite{segal,hwang}. In our model, we find an asymmetry in the two-particle 
current when either the on-site energies in the dot or the couplings of the 
dot with the leads break left-right symmetry. 

Recently we developed a technique employing the Lippman-Schwinger 
scattering theory to study nonequilibrium transport in an 
open system with electron-electron interactions in a localized region 
\cite{dhar}. In this paper we extend that method to investigate quantum 
transport in more realistic models. Compared to our previous study, here we 
incorporate on-site energy in the dot as well as arbitrary tunnelings between 
the dot and the leads. In experiments, the on-site energy in 
the dot is realized through a plunger gate attached to the dot while 
quantum point contacts between the dot and the leads control the tunneling 
strength. We show how the two-electron scattering states and the 
corresponding current can be evaluated for an arbitrary strength of the 
Coulomb interaction. We then use a two-particle scattering approximation
to find the current in the presence of Fermi seas in the leads. 

We study a model of a quantum dot coupled to leads on its left and right 
sides; we first consider spinless electrons for simplicity. The model is 
described by a tight-binding Hamiltonian; the dot consists of two sites 
$(0,1)$ with an interaction $U$ if both sites are occupied by electrons. 
The Hamiltonian is
\bea H &=& H_{LR} ~+~ H_D ~+~ V, \label{ham1} \\
H_{LR} &=& - \sum_{x=-\infty}^{\infty} \hspace{-0.15cm}' ~~
(c_x^\dg c_{x+1} + c_{x+1}^\dg c_x ), \nn \\ 
H_D &=& e_0 n_0 +e_1 n_1 -(c_0^\dg c_1 + c_1^\dg c_0) \nn \\
& & - \g_0 (c^\dg_{-1} c_0 +c_0^\dg c_{-1}) - \g_1 (c^\dg_1 c_2 +c_2^\dg 
c_1), \nn \\
V &=& U n_0 n_1, \nn \eea
where ${\hat n}_x=c_x^\dg c_x$ is the number operator at site $x$, and 
$\sum'$ means summation over all integers omitting $x=-1,0,1$.
Note that we have set the hopping $\g_{x,x+1} = 1$ for all $x$
except $x= -1$ and 1 where it takes the values $\g_0$ and $\g_1$.

The energy of a single particle with wave number $k$ is given by $E_k = - 2 
\cos k$, where $- \pi < k < \pi$. The wave function $\phi_k (x)$ for a 
particle incident on the dot from the left or from the right can be found in 
terms of the dot parameters $e_i$ and $\g_i$. The explicit expressions for 
these wave functions and the reflection and transmission amplitudes are as 
follows. For a particle incident from the left (with $0 < k < \pi$), we have
\bea \phi_k (l) &=& e^{ikl} ~+~ r_k e^{-ikl} ~~{\rm for}~~ l \le -1, \nn \\
&=& (1+r_k)/\g_0 ~~{\rm for}~ l = 0, ~{\rm and}~ t_k e^{ik} /\g_1 ~~ 
{\rm for}~ l = 1, \nn \\
&=& t_k e^{ikl} ~~{\rm for}~~ l \ge 2, \nn \\
t_k &=& \f{-2i\g_0 \g_1 e^{-ik}\sin k}{(e_1-E_k-\g_1^2 e^{ik})
(e_0-E_k-\g_0^2 e^{ik})-1}, \nn \\
r_k &=& \f{1-(e_1-E_k-\g_1^2 e^{ik})(e_0-E_k-\g_0^2 e^{-ik})}{
(e_1-E_k-\g_1^2 e^{ik})(e_0-E_k-\g_1^2 e^{ik})-1}. \eea
For a particle incident from the right (with $-\pi < k < 0$), we have
\bea \phi_k (l) &=& t_k e^{ikl} ~~{\rm for}~~ l \le -1, \nn \\
&=& t_k/\g_0 ~{\rm for}~ l = 0, ~{\rm and}~ (e^{ik}+r_k e^{-ik})/\g_1 ~
{\rm for}~ l = 1, \nn \\
&=& e^{ikl} ~+~ r_k e^{-ikl} ~~{\rm for}~~ l \ge 2, \nn \\
t_k &=& \f{2i\g_0 \g_1 e^{ik}\sin k}{(e_1-E_k-\g_1^2 e^{-ik})
(e_0-E_k-\g_0^2 e^{-ik})-1}, \nn \\
r_k &=& \f{e^{2ik}[1-(e_1-E_k-\g_1^2 e^{ik})(e_0-E_k-\g_0^2 e^{-ik})]}{
(e_1-E_k-\g_1^2 e^{-ik})(e_0-E_k-\g_0^2 e^{-ik})-1}. \eea
We note that the transmission probability $|t_k|^2$ is the same for wave 
numbers $k$ and $-k$;
we will see below that the two-particle current will generally not have 
this symmetry as a result of the interaction. For a weakly coupled dot with 
$\g_i \to 0$, there is a one-particle resonance in the transmission if the 
energy of the incoming particle is given by one of two special values,
\beq E_{1r\pm} ~=~ \f{1}{2} ~[ e_0 + e_1 \pm \sqrt{(e_0 - e_1)^2 + 4}], 
\label{res1} \eeq
provided that the energy lies within the range $[-2,2]$. If the energy lies 
outside the range $[-2,2]$, it corresponds to a bound state rather than a 
transmission resonance. Eq. (\ref{res1}) corresponds to the one-particle 
eigenvalues of the two-site Hamiltonian $e_0 n_0 +e_1 n_1 -(c_0^\dg c_1 + 
c_1^\dg c_0)$.

The two-particle scattering states can be found exactly in this model 
\cite{dhar}. If $H_0 = H_{LR} + H_D$ denotes the noninteracting Hamiltonian,
and $E_k$ and $\phi_k (x)$ are the one-particle energies and wave functions,
the noninteracting two-particle energies and wave functions are given by
$E_{\bk}=E_{k_1}+E_{k_2}$ and $\phi_{\bk} (\bx) = \phi_{k_1} (x_1) 
\phi_{k_2} (x_2) - \phi_{k_1} (x_2) \phi_{k_2} (x_1)$, where $\bk=(k_1,k_2)$
and $\bx=(x_1,x_2)$. A scattering eigenstate of the total Hamiltonian $H = H_0
+ V$ is then given by the Lippman-Schwinger equation $|\psi\ra = |\phi\ra 
+ G_0^+(E) V |\psi\ra$, where $G_0^+(E) = {1}/{(E- H_0 +i \e)}$. In the 
position basis $|\bx \ra$, we obtain
$\psi_\bk (\bx) ~=~ \phi_\bk (\bx) ~+~ UK_{E_\bk} (\bx)~ \psi_\bk (\b0)$,
where $\b0 \equiv (0,1)$, $K_{E_\bk}(\bx) = \la \bx | G_0^+(E_\bk) |\b0 
\ra$ has the explicit form 
\beq K_{E_\bk}(\bx) ~=~ \f{1}{2} \int_{-\pi}^\pi \int_{-\pi}^\pi \f{dq_1 
dq_2}{(2 \pi)^2} \f{\phi_{\bq} (\bx) \phi_{\bq}^* (\b0)}{E_\bk - E_\bq + i\e}~,
\eeq
and $\psi_\bk (\b0) = \phi_\bk (\b0)/[1 - U K_{E_\bk} (\b0)]$. 
Using this approach, we find that two particles incident with wave
numbers $k_1,k_2$ scatter to a continuous range of final wave numbers
$q_1,q_2$. This is because the interaction breaks translation invariance;
hence the total momentum is not conserved although the energy is. This 
suggests that the model is not solvable by the Bethe ansatz \cite{dhar}.

We now evaluate the two-particle current through the dot; this is given by
the expectation value of the operator 
\beq \hj ~=~ -i ~\g_{x,x+1} (c^{\dg}_x c_{x+1} - c^{\dg}_{x+1} c_x), \eeq
in the scattering state $|\psi_{\bf k} \ra = |\phi_\bk \ra+| S_\bk \ra$, where
$|S_\bk \ra \equiv G_0^+(E) V |\psi_\bk \ra$ is the interaction induced 
correction to the scattering state. Since $[{\hat n}_x , H] = i({\hat j}_{x-1}
- \hj)$, $\la \hj \ra$ is independent of $x$ in any eigenstate
of $H$. Let us write $\la \hj \ra = j_I + j_C + j_S$, where $j_I = \la 
\phi_\bk | \hj |\phi_\bk \ra$, $j_C = \la \phi_\bk | \hj | S_\bk \ra + \la 
S_\bk | \hj | \phi_\bk \ra$, and $j_S = \la S_\bk | \hj | S_\bk \ra$. We will 
now calculate all these terms. If we assume that the
system has $\cal N$ sites, we find that $j_I = 2 {\cal N} (\sin k_1 |t_{k_1}|^2
+ \sin k_2 |t_{k_2}|^2)$. Next, $j_C = 2~ {\rm Im} ~\la \phi_\bk | (c^\dg_x 
c_{x+1} - c^\dg_{x+1} c_x) |S_\bk \ra$, and
\bea & & \la \phi_\bk| c^\dg_{x_1} c_{x_2} |S_\bk\ra = \f{\phi_\bk ({\b0})}{
1/U- K_{E_\bk}({\b0})} ~\int^\pi_{-\pi} \f{dq}{2\pi} ~\phi_q (x_2) \nn \\
& & ~~\times ~~\Big( \f{\phi^*_{k_2}(x_1) \phi^*_{k_1 q}({\b0})}{E_{k_2} -E_q
+ i\e} - \f{\phi^*_{k_1}(x_1) \phi^*_{k_2 q}({\b0})}{E_{k_1}- E_q +i\e} \Big).
\label{jc} \eea
Finally, $j_S = 2~{\rm Im} ~\la S_{\bf k}|c^\dg_x c_{x+1}|S_{\bf k}\ra$, and
\bea \la S_\bk|c^\dg_x c_{x+1}|S_\bk\ra = \f{|\phi_\bk( {\b0})|^2}{|1/U-K_{
E_\bk}({\b0})|^2} \int_{-\pi}^\pi \f{dq}{2\pi} I_0(q) I_1^*(q), \nn \\
{\rm where}~I_s (q) = \int_{-\pi}^{\pi} \f{dq_1}{2\pi} 
\f{\phi_{qq_1}(\b0) \phi^*_{q_1}(x+s)}{E_\bk -E_{qq_1} -i\e},~~s=0,1. 
\label{js} \eea
For a small interaction strength $U$, we see that $j_C$ and $j_S$ are generally
of order $U$and $U^2$ respectively. On the other hand, they have non-zero and 
finite limits when $U \to \infty$. We can use Eqs. (\ref{jc}-\ref{js}) to 
compute $\la \hj \ra$ at any convenient value of $x$. (The extra factor of 
$\cal N$ that $j_I$ has with respect to $j_C$ and $j_S$ will disappear when 
we consider the thermodynamic limit below).

\begin{figure}[t] \includegraphics[width=3.2in]{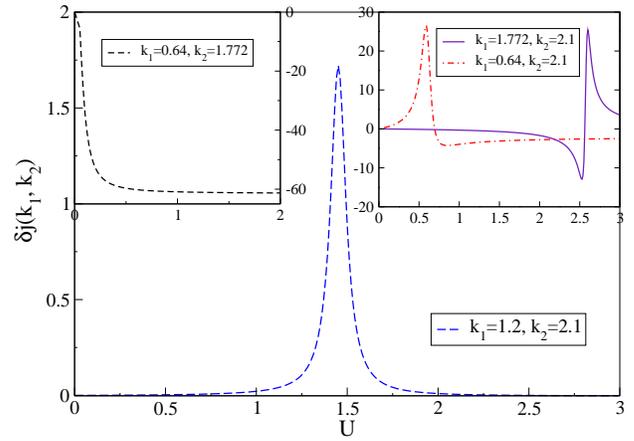}
\caption{(Color online) Plots of $\dj(k_1,k_2)$ versus $U$, for $e_0 = e_1 = 
-0.6$, $\g_0 = \g_1 = 0.2$. Right and left insets show plots of $\dj 
(k_1,k_2)$ versus $U$ when one or both of the incident energies correspond 
to one-particle resonances for the same parameter set.} \label{twopleres1} 
\end{figure}

We have used Eqs. (\ref{jc}-\ref{js}) to numerically compute the correction to
the current $\dj (k_1,k_2) \equiv j_C + j_S$ caused by the interaction. 
[In the numerical calculations, the integrals were approximated by summations 
with a small grid size $dq$ and several small values of $\e$ satisfying 
$dq \ll \e \ll 1$. The results were then linearly extrapolated to the limit 
$\e \to 0$.] We discover two interesting phenomena:

\noi (i) First, we find that $\dj (k_1,k_2)$ as a function of $U$ has peaks 
at certain values of the energies of the two incident states. We will call 
this an interaction induced two-particle resonance; this was recently
noticed in Ref. \cite{lebedev}. To 
understand this, let us first set the dot-lead couplings $\g_i = 0$. In that 
case a state in which sites $0$ and $1$ are occupied by one particle each is 
an eigenstate of $H_0$ with energy $e_0 + e_1$, and of $H$ with energy $e_0 
+ e_1 + U$. Then $K_{E_{\bf k}}({\b0}) = \la \b0 | 1/(E_\bk - H_0 + i \e)| 
\b0 \ra$ will be purely real and equal to $1/(E_\bk - e_0 - e_1)$ if $E_\bk 
\ne e_0 + e_1$. We now turn on small values of the $\g_i$, 
and consider two particles coming from the leads with a total energy $E_\bk = 
E_{k_1} + E_{k_2}$, where $E_{k_i}$ are {\it not} at the one-particle 
resonance energies $E_{1r\pm}$, so that $j_I$ is close to 0. We expect that 
if $E_\bk \ne e_0 + e_1$, the real and imaginary parts of 
$K_{E_{\bf k}}({\b0})$ will remain close to $1/(E_\bk - e_0 - e_1)$ and 0 
respectively. It is now clear from the pre-factors in the expressions in 
Eqs. (\ref{jc}-\ref{js}) that $\dj(k_1,k_2)$ will show a peak, as a function 
of $U$, at $1/U-K_{E_{\bf k}}({\b0}) = 0$, i.e., at $E_\bk = E_{2r}$, where 
the two-particle resonance energy is given by
\beq E_{2r} ~=~ e_0 ~+~ e_1 ~+~ U. \label{res2} \eeq
Fig. \ref{twopleres1} illustrates the effects of two-particle resonance. The 
main plot shows a peak in $\dj(k_1,k_2)$ at $U \simeq 1.45$ compared to 
$U = 1.48$ expected from Eq. (\ref{res2}); the deviation is presumably 
due to the small but finite values of $\g_0$ and $\g_1$. The right inset 
shows what happens when one of the incident energies is at a one-particle 
resonance; then the two-particle resonance, occurring at $U = 2.6$ for 
$(k_1,k_2)=(1.772,2.1)$ and $U = 0.6$ for $(k_1,k_2)= (0.64,2.1)$, produces 
a rapid variation in the current with $U$ due to the denominator $1/U- 
K_{E_\bk}({\b0})$ in Eq. (\ref{jc}) going through zero. The left inset of 
Fig. 1 shows what happens when both the incident energies correspond to 
one-particle resonances; the interaction causes backscattering and suppresses 
the one-particle resonance by a large amount because the pre-factor of 
$\phi_\bk ({\b0})$ in Eqs. (\ref{jc}-\ref{js}) is large for one-particle 
resonances. 

\vskip .7 true cm
\vskip .3 true cm

\noi (ii) Secondly, we find that $\dj (k_1, k_2) \ne - \dj (-k_1,-k_2)$ if the 
system is not invariant under the parity transformation $x \leftrightarrow 
1-x$, i.e., if either $e_0 \ne e_1$ or $\g_0 \ne \g_1$. The reason for current
asymmetry is the re-distribution of the electrons' momentum after scattering 
from the dot along with the absence of spatial inversion symmetry in the 
model. It can be understood quantitatively if $\g_0$ and $\g_1$ are both 
small but differ greatly in magnitude, and if $k_1,k_2$ have the same sign. 
We see from Eqs. (\ref{jc}-\ref{js}) that the strength of the interaction 
depends on the 
probability $|\phi_\bk (\b0)|^2$ of finding the two particles at sites 0 and 
1. If both the particles come from the left (right) lead, their joint 
amplitude of reaching sites 0 and 1 is proportional to $\g_0^2$ ($\g_1^2$). 
Hence, $|\phi_\bk (\b0)|^2$ will be proportional to $\g_0^4$ ($\g_1^4$) if
$k_1, k_2 > 0$ ($< 0$); hence $\dj$ will be quite different in the two cases 
if $\g_0$ and $\g_1$ have very different values. For instance, if $e_0 = -
0.8$, $e_1 = -0.3$, $\g_0 = 0.1$, $\g_1 = 0.3$, $U=1$, $k_1 = 1$ and $k_2 = 
2$, we find numerically that $\dj (k_1,k_2) = 0.031$ and $\dj (-k_1,-k_2) = 
-1.014$. We note that the ratio $|\dj (-k_1,-k_2)/\dj (k_1,k_2)| \simeq 33$
which is of the same order of magnitude as $\g_1^4 /\g_0^4 = 81$.

We now examine whether the two-particle resonance
remains visible when we consider a many-electron system. Let us compute the 
current when the left (right) leads are at zero temperature and chemical 
potentials $\mu_L$ ($\mu_R$). This requires us to find 
$N$-particle scattering states and then take the limit $N \to \infty$. It is 
difficult to find such states exactly in our model. We therefore make the 
approximation of considering only two-particle scattering \cite{dhar}; this 
is justified if either the density is so low that three-electron scattering 
can be ignored \cite{rech}, or if $U \ll 2 \pi \sin k_F /k_F$. [The latter 
condition arises as follows. In the simple case with $e_0 = e_1 = 0$ and 
$\g_0 = \g_1 = 1$, the interaction $V$ in Eq. (\ref{ham1}) can be written in a
Hartree-Fock approximation as $U (\la n_0 \ra n_1 + \la n_1 \ra n_0)$, where 
the mean density is related to the Fermi momentum as $\la n_i \ra = 
k_F /\pi$. At the Fermi momentum $k_F$, the reflection probability for this 
one-particle problem is much less than 1 if $U \la n_i \ra$ is much less
than the Fermi velocity $2 \sin k_F$. We thus require that $U \ll 2\pi \sin 
k_F /k_F$.] Within the two-particle approximation, 
we write $|\psi_{\bkn} \ra= |\phi_{\bkn} \ra + |S_{\bkn}\ra$, where the 
amplitude of scattering from a wave vector $\bkn = \{k_1k_2...k_N\}$ to 
a wave vector $\bqn= \{q_1q_2...q_N\}$ is given by
\bea \la \bqn| S_{\bkn}\ra &=& \sum_{\bq_{2} \bk_2} (-1)^{P+P'} \la \bq_2|
S_{\bk_2} \ra \la \bq'_{N-2}{|\bk'_{N-2}}\ra, \nn \\
\la \bq_{2}| S_{\bk_2}\ra &=& \frac{\phi_{\bq_2}^*(\b0) 
\phi_{\bk_2} (\b0)}{(1/U - K_{E_{\bk_2}} ({\b0})) (E_{\bk_2}-E_{\bq_2}+i \e)},
\label{npart} \eea
where $\bq_2$ ($\bk_2$) denotes a pair of momenta chosen from the set $\bqn$ 
($\bkn$), $\bq_{N-2}'$ ($\bk_{N-2}'$) denotes the remaining $N-2$ momenta, and
$P$ ($P'$) is the appropriate number of permutations. Using Eq. (\ref{npart}),
we can calculate the current expectation value for the state $|\psi_{\bkn}
\ra$. The noninteracting current is $j_I = 2 
{\cal N}^{N-1} \sum_{j=1}^N \sin k_j |t_{k_j}|^2$. The correct normalization
is obtained by dividing by a factor of ${\cal N}^N$; in the thermodynamic limit
$N, {\cal N} \to \infty$, this gives $j_I = \int_{k_R}^{k_L} (dk/2\pi) 2 \sin
k |t_k|^2$. Here $-k_R$ ($k_L$) is the Fermi wave number of the right (left) 
lead lying in the range $[-\pi,0]$ ($[0,\pi]$); it is related to the 
corresponding chemical potentials by $\mu_{R/L} = - 2 \cos k_{R/L}$. Inserting
factors of $\hbar$ and the charge $e$, the above expression for $j_I$ gives 
the current for the noninteracting system to be $I = (e/h) 
\int_{\mu_R}^{\mu_L} dE |t_k|^2$, where $E=-2 \cos k$. We now compute the
correction to this current, $\dj_N$, caused by the interaction. Using the
normalization given above, we find that $\dj_N= (1/2{\cal N}^2) \sum_{r,s} 
\dj( k_r,k_s)$; in the thermodynamic limit, this gives the correction to be
\bea \dj ~=~ \f{1}{2} \int_{-k_R}^{k_L} \int_{-k_R}^{k_L} \f{dk_1 d k_2}{
(2 \pi)^2} ~\dj (k_1, k_2). \label{deljn1} \eea
We know that $\dj = 0$ if there is no voltage bias, i.e.,
if $k_R = k_L$. Hence, if $k_R < k_L$, Eq. (\ref{deljn1}) reduces to 
\bea \dj = \Big[ \int_{k_R}^{k_L} \int_{-k_R}^{k_R} + \f{1}{2} 
\int_{k_R}^{k_L} \int_{k_R}^{k_L} \Bigr] \f{dk_1 d k_2}{(2 \pi)^2} \dj 
(k_1, k_2). \label{deljn2} \eea
In the zero bias limit $\mu_R \to \mu_L$ ($k_R \to k_L$), the contributions 
of the two integrals in Eq. (\ref{deljn2}) are of order $|\mu_R - \mu_L|$ and 
$|\mu_R - \mu_L|^2$ respectively.

Now we study whether the two-particle resonance remains observable after 
doing the $k_1,k_2$ integrals in Eq. (\ref{deljn2}). This is shown 
in Fig. 2 where the dot parameters are the same as in Fig. 1, and the average 
chemical potential $\mu_0 = (\mu_L + \mu_R)/2$ is kept fixed at $0.95$.
The main plot shows peaks in a plot of the total current $j = j_I 
+ \dj$ versus $U$; the reason for these peaks is the following. Since the 
bias $\Delta \mu = \mu_L - \mu_R$ is small, the first integral in Eq. 
(\ref{deljn2}) dominates; hence the variable $k_1$ stays close to 
$k_0 = 2.07$ corresponding to the energy $E_1 = 0.95$. The other variable 
$k_2$ goes over a range of about $[-2.07,2.07]$; the corresponding range for 
$E_2$, $[-2,0.95]$, includes the {\it one-particle} resonance energies 
given in Eq. (\ref{res1}), $E_{1r\pm} = -1.6$ and $0.4$, where there is 
a high probability for this particle to enter the dot. When the
two-particle energy $E_1 + E_2 = -0.65$ or $1.35$ happens to be equal to the
two-particle resonance energy $e_0 + e_1 + U$, we get a large contribution
to $\dj$. This predicts the peaks to lie at $U = 0.55$ and $2.55$ which are
close to the values of $0.53$ and $2.52$ observed in Fig. 2. We also 
note that for the three values of the bias $\Delta \mu = 0.02, ~0.04, ~0.08$,
the values of $j$ at the peaks lie in the range $1 - 6 \times 10^{-3}$ which 
is much larger than the interaction-independent current $j_I$ which lies in
the range $1 - 4 \times 10^{-5}$. We emphasize that the two-electron resonance
occurs near a chemical potential ($0.95$) which lies well above the 
one-particle resonance energies $E_{1r\pm}$; thus an electron at the chemical 
potential transmits through the dot only due to the interaction $U$.
The inset of Fig. 2 shows the current versus the bias for $U=0.52$ which 
corresponds to the first peak in the main figure, and $U=1.1$ which lies
between the peaks; we see that the conductance is much larger in the first 
case. In all our calculations, we have ensured that the bias is not large 
enough for either of the chemical potentials to lie close to a one-particle
resonance; otherwise the two-particle resonance might get masked by a 
one-particle resonance.

The analysis in this paper can be readily extended to the case of spin-1/2
electrons. We consider a simple model of a dot consisting of only one site (at
$x=0$) where there is an on-site energy $e_0$ and an interaction of the form 
$U n_{0\ua} n_{0\da}$. This can lead to scattering between two electrons in 
the singlet channel but not in the triplet channel. The scattering and the 
resultant correction to the current can again be studied using the 
Lippman-Schwinger formalism. We again find that a two-electron resonance
can occur at an energy given by $2e_0 + U$ if the dot-lead couplings are 
small. In addition to this, the interaction can now also lead 
to spin entanglement \cite{oliver}. Namely, if a spin-up and a 
spin-down electron are incident on the dot in a spin-uncorrelated state with
a total energy which is equal to the two-particle resonance energy, 
the two electrons will emerge in a singlet state after scattering.

To summarize, we have studied a model of a quantum dot which is a small
region in which electrons interact. The scattering of two particles due to 
the interaction is studied exactly. We find that a two-particle resonance 
occurs if the incident energies and the dot parameters satisfy a certain 
relation. Further, the interaction generally leads to an asymmetry in 
the current if the incident wave numbers are reversed; for a many-electron
system with no inversion symmetry and strong Coulomb interactions, the
current asymmetry can be shown by using a master equation approach
\cite{bagrets}. We then use a two-electron 
perturbative approach to show that the two-particle resonance can survive 
for the many-electron system which arises when the leads are Fermi seas 
with certain chemical potentials; the resonance occurs if the dot parameters 
($e_i, \g_i, U$) and the chemical potentials are related in a particular way, 
and the resultant current can be much larger than $j_I$. These phenomena can
persist if we consider a more realistic model of a dot which has interactions 
over a larger region.
It would be interesting to look for these effects experimentally in quantum 
dot systems.

\vskip .7 true cm
\begin{figure}[t] \includegraphics[width=3.2in]{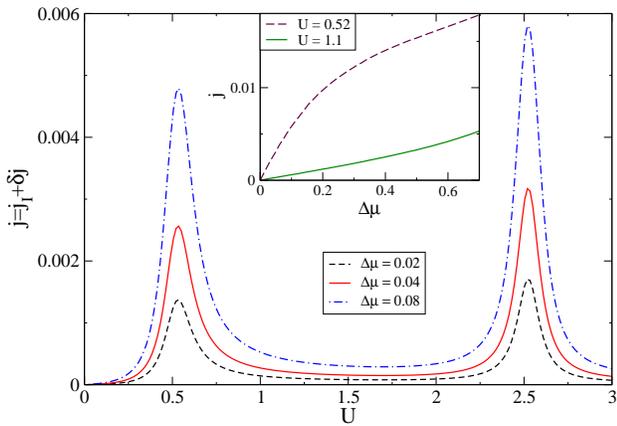}
\caption{(Color online) Plots of total current $j = j_I + \dj$ versus $U$, 
for $e_0 = e_1 = -0.6$, $\g_0 = \g_1 = 0.2$, and bias $= 0.02, ~0.04, ~0.08$.
Inset shows $j$ versus bias for $U=0.52$ and $1.1$.} \label{twopleres2} 
\end{figure}
\vskip .3 true cm

We thank M. B\"uttiker, Y. Imry, D. E. Logan, A. Nitzan, S. Rao, B. Sriram 
Shastry and E. V. Sukhorukov for fruitful discussions. D.S. thanks DST,
India for financial support under Project No. SR/S2/CMP-27/2006.

\end{document}